\documentstyle[12pt,a4,epsf]{article}
\newcommand{\nn}{\nonumber}

\newcommand{\be}{\begin{equation}}
\newcommand{\ee}{\end{equation}}
\newcommand{\bea}{\begin{eqnarray}}
\newcommand{\eea}{\end{eqnarray}}
\newcommand{\ag}{{\alpha_g}}
\newcommand{\ay}{{\alpha_y}}
\newcommand{\ak}{{\alpha_{\kappa}}}

\begin{document}

\begin{flushright}
KUNS-1948\\ 
KANAZAWA-04-19
\end{flushright}
\vspace{10mm}
\begin{center}
{\Large \bf
Democratic mass matrices induced by strong gauge
\vspace{3mm}\\
dynamics and large mixing angles for leptons
}
\vspace{10mm}\\
Tatsuo Kobayashi$^{a}$\footnote{
E-mail: kobayash@gauge.scphys.kyoto-u.ac.jp
},
Hiroaki Shirano$^{b}$\footnote{
E-mail: shirano@hep.s.kanazawa-u.ac.jp
}
and Haruhiko Terao$^{b}$\footnote{
E-mail: terao@hep.s.kanazawa-u.ac.jp
}
\vspace{5mm}\\
$^a$Departmant of Physics, Kyoto University, \\
Kyoto 606, Japan 
\vspace{3mm}\\
$^b$Institute for Theoretical Physics, Kanazawa
University, \\
Kanazawa 920-1192, Japan
\vspace{10mm}\\
{\large Abstract}
\vspace{5mm}\\
\begin{minipage}{130mm}{
We consider dynamical realization of the democratic type Yukawa coupling
matrices as the Pendelton-Ross infrared fixed points.
Such fixed points of the Yukawa couplings
become possible by introducing many Higgs fields, which are
made superheavy but one massless mode.
Explicitly, we consider a strongly coupled GUT based on 
$SU(5) \times SU(5)$, where rapid convergence to the infrared
fixed point generates sufficiently large mass hierarchy for
quarks and leptons.
Especially, it is found that the remarkable difference between mixing 
angles in the quark and lepton sectors may be
explained as a simple dynamical consequence.
We also discuss a possible scenario leading to the realistic mass 
spectra and mixing angles for quarks and leptons.
In this scheme, the Yukawa couplings not only for top but 
also for bottom appear close to their quasi-fixed points 
at low energy and, therefore, $\tan \beta$ should be large.
}
\end{minipage}
\end{center}

\newpage
%%%%%%%%%%%%%%%%%%%%%%%%%%
\section{Introduction}

Yukawa couplings are not direct observables, since
they change depending on the choice of flavor basis.
What we know in principle are the hierarchical masses of quarks 
and leptons, and the mixing matrices.
The recent neutrino oscillation experiments 
\cite{SK,SNO,KM,K2K} indicate strongly that
neutrinos are massive and lepton flavors are mixed.
It is remarkable that the mixing angles 
($\sin^2 2 \theta_{\rm sun} \sim 0.84$ and
$\sin^2 2 \theta_{\rm atm} \sim 1.0$) are as large as 
nearly bi-maximal, which are
in sharp contrast with the small mixing
angles among quarks.

There have been proposed various phenomenological models 
explaining origin of the Yukawa couplings accommodating
these features by assuming certain flavor symmetries. 
\cite{models}.
The most popular one would be the Froggatt-Nielsen
mechanism \cite{FN} based on abelian flavor symmetries.
In this line of thought, hierarchical structures
among the Yukawa couplings are nicely explained.
However the abelian symmetry itself cannot fix 
O(1) ambiguity of parameters. Therefore
structure of the mixing angles in the lepton sector
as well as the neutrino masses are explained no more
than that those parameters are not  hierarchical.
Also the flavor differences are just input in the
abelian flavor charges apriori.

Contrary to this, non-abelian flavor symmetries can
impose firm relations among some elements of the
Yukawa matrices. Therefore various non-abelian symmetries
have been considered to explain the mixing angles 
in the lepton sector as well as the hierarchical
masses simultaneously \cite{SU3}.
The democratic Yukawa matrices have been an attractive
possibility of this kind 
\cite{quarksector,leptonsector,S3,GUT,S3improve}. 
The rigid democratic matrix,
in which all elements are identical
\footnote{Phenomenologically the so-called extended democratic 
mass matrices \cite{extended} are equally considerable.
However we restrict ourselves to the rigid cases in this paper.}
, may be explained
by assuming {\it e.g.} $S_3$ flavor symmetries
for three generations of quarks and leptons.
It is amusing that one of the mass eigenvalues appears much
larger than others, though there is no apriori difference
among flavors.
The $S_3$ flavor symmetry is supposed to be broken slightly
to give rise to small masses for the first and second generations.
It has been known for some time that Fritzsch type mass matrices
for quarks are realized after introducing a simple form of 
$S_3$ breaking parameters \cite{quarksector}.
Moreover the lepton sector mixing angles uncovered by
neutrino oscillation experiments may be explained in the democratic
framework rather well \cite{leptonsector,S3}.

However no comprehensive explanation has been given to the
origin of the flavor symmetry breaking terms 
in this framework.
Moreover the $S_3$ flavor symmetry allows the mass matrix 
proportional to an identity matrix other than the democratic
type matrix for the neutrino sector.
The mixing matrix for the lepton sector can be obtained,
only if we assume the neutrino mass matrix to be almost
diagonal. Namely the democratic part must be small
with some other reasons than the flavor symmetry 
\cite{leptonsector,S3}.
The same problem remains in the see-saw context, where
both of Yukawa couplings and mass matrix of the 
right-handed neutrinos should be almost diagonal.

On the other hand, we face with a more stringent
problem in considering the grand unified theories.
In {\it e.g.} the SU(5) GUT, the matters belong to
a ${\bf 10}$ and a $\bar{\bf 5}$.
However $S_3$ symmetry allows an identity matrix for the
Yukawa couplings among ${\bf 10}$'s, which must be
strictly forbidden for the quark mass hierarchy \cite{GUT}.
Thus the phenomenologically postulated forms for the
democratic Yukawa matrices are not explained only
by flavor symmetries.

Indeed it is not a unique way to introduce some flavor 
symmetries and their small breakings in order to
explain hierarchy among various couplings.
For example, large mass hierarchy may be generated by
overlapping of the wave functions in extra dimensions.
Renormalization with large anomalous dimensions induced
by strong dynamics also can generate mass hierarchy \cite{NS}.
Similarly power-law running of the Yukawa couplings 
in the extra dimensions has been also utilized \cite{powerlaw}.

In this paper we consider the models in which the democratic type
of Yukawa matrices are realized as infrared attractive fixed 
point couplings \cite{PR,Ross,AK,powerlaw}.
It has been known for some time that ratio of the Yukawa couplings
and the gauge coupling approach rapidly to the so-called
Pendelton-Ross fixed points (PRFP) towards infrared
\cite{PR}.
If all elements of the Yukawa coupling matrix have an identical
PRFP, then the democratic type matrix may be achieved
dynamically. In order to make this possible, we need to introduce
many Higgs fields. 
Also a special kind of mass terms for the multi Higgs fields are 
assumed so that they obtain masses of the GUT scale except
for one massless mode, which is identified with Higgs in the
low energy theory.

Indeed this idea is not new. 
Abel and King\cite{AK} have already considered several years ago
except for the neutrino sector.
In this paper, however, we stress that this mechanism is
free from the difficulty in realizing democratic Yukawa
matrices by the flavor symmetry.
Moreover we are going to show that the typical difference 
in the mixing angles between quark and lepton sectors may be
explained as a simple dynamical consequence in this framework.
Our mechanism predicts very small neutrino Yukawa couplings, and
therefore relatively light right handed neutrinos in the
see-saw models.

Explicitly, we treat a supersymmetric
GUT model based on $SU(5) \times SU(5)$ gauge group,
in which the doublet-triplet problem is solved nicely \cite{DT,witten}.
There one $SU(5)$ gauge interaction coupled with quarks and 
leptons is assumed to be strong and another is weak just as
in the conventional SU(5) GUT.
Then renormalization group (RG) from the Planck scale to
the GUT scale shows that the Yukawa couplings in the GUT model 
are aligned in a very good accuracy due to strong dynamics.
After diagonalizing the Yukawa matrix, only one coupling is
found to be much larger than others.
Further we will show that the realistic Yukawa  coupling matrices
may be obtained within a simple setup.

In this kind of scenarios, all of the Yukawa couplings, except
for the neutrino sector, are given to be quite large at high
energy.
Because the Yukawa coupling and the gauge
coupling are of the same order at the PRFP. 
However heavy top quark mass (178GeV) indicates
that top Yukawa coupling is close to the so-called
quasi-fixed point \cite{quasifp}.
This means that top Yukawa coupling can be fairly large
at the GUT scale, which is a favorable point of our
scenario.
However Yukawa couplings not only of top quark but also of 
bottom quark and tau lepton are all large. 
Therefore large $\tan \beta$
is predicted in this kind of scenarios.

The article is organized as follows.
In section 2 we give a brief summary of the democratic
type of mass matrices and the $S_3$ flavor symmetry.
In section 3 we present the general ideas leading to the
democratic Yukawa matrices at low energy by strongly
coupled gauge dynamics.
There also the superpotential for the multi Higgs 
fields and their mass spectra are discussed. 
In section 4 we give an explicit model based on the 
$SU(5) \times SU(5)$ GUT. Then it is seen that
sufficiently large hierarchy in couplings can be 
generated by RG from the Planck scale to the 
GUT scale.
We also discuss predictions of this kind of scenarios
and consistency with particle masses.
There we work out the Yukawa couplings for top, bottom and tau
obtained at low energy scale in the explicit model.
The neutrino Yukawa couplings are discussed in section 5.
There it is shown that the large mixing angles among
leptons may be realized as a dynamical consequence, if 
the right-handed neutrinos have new Yukawa couplings 
to some other fields. 
In section 6 we consider minute structure of the
democratic mass matrices producing realistic masses
and mixing angles. 
Finally we devote section 7 to conclusions and 
discussions including comments
on the extra dimensional setup and also on the soft 
supersymmetry breaking parameters.

\section{Democratic mass matrices and $S_3$ flavor symmetry}
In this section we give a brief review of the democratic
mass matrices for quarks and leptons from the view point
of $S_3$ flavor symmetry.
The democratic mass matrices give a
phenomenologically successful description. 
However it is found to be necessary to constrain the 
parameters allowed by the symmetry further.
Namely the $S_3$ symmetry is not sufficient to give
viable mass matrices.
Indeed some attempts have been done in order to 
give comprehensive grounds to the viable democratic mass
matrices so far \cite{S3improve}.

The rigid democratic matrix $J$ is diagonalized as  
\be
J  =  \frac{1}{3}
\left(
\begin{array}{ccc}
1 & 1 & 1 \\
1 & 1 & 1 \\
1 & 1 & 1
\end{array}
\right)
= A
\left(
\begin{array}{ccc}
0 & 0 & 0 \\
0 & 0 & 0 \\
0 & 0 & 1
\end{array}
\right)
A^T,
\ee
by the diagonalization matrix
\be
A = 
\left(
\begin{array}{ccc}
1/\sqrt{2} & 1/\sqrt{6} & 1/\sqrt{3} \\
-1/\sqrt{2} & 1/\sqrt{6} & 1/\sqrt{3} \\
0 & -2/\sqrt{6} & 1/\sqrt{3}
\end{array}
\right).
\label{A-matrix}
\ee
For the democratic mass matrix, two vanishing eigenvalues
are regarded as masses for the first and the second generations
in the first approximation.
Their masses are given by small deviations from the rigid matrix.
It would be said that the interesting feature of this approach
is the idea of flavor democracy, namely that there is no 
difference between flavors apriori.

In the standard model (SM), 
the democratic mass matrices are realized by assuming
$S_3$ flavor symmetry in 
each kind of the SM matter fields.
Hereafter we consider the minimal supersymmetric case, 
the MSSM. The superpotential of the MSSM is given as
\be
W = Y^u_{ij} ~Q_i u_j H^u   +  Y^d_{ij} ~Q_i d_j H^d 
 +  Y^e_{ij} ~L_i e_j H^d  
+  \frac{\kappa_{ij}}{2 M_R} ~L_i L_j H^u H^u,
\label{MSSM}
\ee
where $i,j = (1, 2, 3)$ represent generations.
Each of $Q_i, u_i, d_i, L_i, e_i$ is assigned 
to a three dimensional representation of a
distinct $S_3$.
Then the Lagrangian is invariant under permutation
of any set of the matter fields.

Let us start with mass matrices for quarks.
It has been known that the realistic mass matrices
are obtained by introducing
small breaking parameters 
$\epsilon_{u(d)} \ll \delta_{u(d)} \ll 1$
of the flavor symmetry as follows \cite{quarksector}.
\bea
M_q & \propto & ~J ~+~
\left(
\begin{array}{ccc}
-\epsilon_q & 0 & 0 \\
0 & \epsilon_q & 0 \\
0 & 0 & \delta_q 
\end{array}
\right) \nn \\
& = &
A
\left(
\begin{array}{ccc}
0 & -\sqrt{1/3}\epsilon_q & -\sqrt{2/3} \epsilon_q \\
-\sqrt{1/3}\epsilon_q & (2/3)\delta_q & -(\sqrt{2}/3) \delta_q \\
-\sqrt{2/3} \epsilon_q & -(\sqrt{2}/3) \delta_q & 1 +(1/3)\delta_q
\end{array}
\right) A^T,
\eea
where $q=u, d$. The matrix in the 2nd line is known as the
Fritzsch type, which offers us a  phenomenologically
good representation.

The mass eigenvalues of these matrices are found to be
\be 
m^q_1 : m^q_2 : m^q_3  \sim 
\frac{\epsilon_q^2}{\delta_q} : \delta_q : 1.
\ee
Therefore ratio of quark masses can be represented by assuming
the breaking parameters of the following order;
\be
\begin{array}{ll}
\epsilon_u \sim 10^{-4}, & \delta_u \sim 10^{-2}, \\
\epsilon_d \sim 10^{-2}, & \delta_d \sim 10^{-1}.
\end{array}
\label{breakings}
\ee
The absolute values of the top and bottom quark masses
are not determined by the flavor symmetry and, therefore,
must be parameterized phenomenologically.
In our dynamical mechanism, we can predict these absolute
values by evaluating the Yukawa couplings at low energy.
This will be discussed later.

The mixing angles turn out to be small, since
the diagonalization matrix for $M_u$ and
$M_d$ are both close to $A$.
Quantitatively the CKM matrix is given in a 
good approximation by
\be
U_{\rm CKM} \sim 
\left(
\begin{array}{ccc}
1 & O(\epsilon_d/\delta_d) & O(\epsilon_d) \\
O(\epsilon_d/\delta_d) & 1 & O(\delta_d) \\
O(\epsilon_d) & O(\delta_d) & 1
\end{array}
\right).
\ee
It is found that the above choice of parameters 
may reproduce the realistic mixing angles also.

Next let us go into the lepton sector.
The mass matrix for the charged leptons are given in a 
similar fashion;
\be
M_e  \propto J ~+~
\left(
\begin{array}{ccc}
-\epsilon_e & 0 & 0 \\
0 & \epsilon_e & 0 \\
0 & 0 & \delta_e 
\end{array}
\right).
\ee
Here, if we take the flavor symmetry breaking parameters as 
\be
\epsilon_e \sim 10^{-2},~~~\delta_e \sim 10^{-1},
\ee
then ratio of the charged lepton masses are parameterized well.
Again the absolute value is undetermined in this approach, 
though it will be determined in our scheme.

However situation for the neutrino mass matrix is somewhat
different.
The $S_3$ flavor symmetry allow the couplings $\kappa_{ij}$
given by Eq.~(\ref{MSSM}) proportional to an identity also.
Therefore the mass matrix is given as
\be
M_{\nu} \propto ~I~+ ~r J
+
\left(
\begin{array}{ccc}
0 & 0 & 0 \\
0 & \epsilon_{\nu} & 0 \\
0 & 0 & \delta_{\nu} 
\end{array}
\right),
\label{neu-mass}
\ee
where $I$ denotes an identity matrix.

It is noted that this mass matrix may explain the
mixing angles in the lepton sector nicely, if
the parameter $r$ in Eq.~(\ref{neu-mass}) is
much smaller than 1.
For vanishing $r$, the lepton mixing matrix (or
the MNS matrix) is just given by $A^T$ (matrix $A$
is given explicitly in Eq.~(\ref{A-matrix})).
Therefore we obtain the three mixing angles as
\be
\sin^2 2\theta_{12} \sim 1,~~~
\sin^2 2\theta_{23} \sim 0.94,~~~
U_{e3} = 0.
\ee
This should be compared with observation
by the recent neutrino experiments \cite{SK,SNO,KM,K2K},
which tells us
\be
\sin^2 2 \theta_{12} \sim 0.84,~~~
\sin^2 2 \theta_{23} \sim 1.0,~~~
|U_{e3}| <   0.23.
\ee
Thus the lepton mixing angles are found to be 
almost explained already without parameter 
$r$ in Eq.~(\ref{neu-mass}).
In other words, a nearly diagonal neutrino mass matrix
is favorable including the flavor symmetry breaking
part from phenomenological point of view.
However we do not find any reasons either why $r$ is so small,
or why the flavor symmetry breaking parameters are input only
at diagonal elements.

This fine-tuning problem remains or becomes more curious 
in the see-saw mechanism.
The superpotential for the neutrino sector is given by
\be
W = Y^{\nu}_{ij} L_i \nu_j H^u ~+~
\frac{1}{2}M_{R ij} \nu_i \nu_j,
\ee
where $\nu_i$ denote the right-handed neutrinos.
Here we need to assume that these right-handed 
neutrinos belong to the same representation 
of $S_3$ as the lepton doublets $L_i$.
If we assume
a distinct $S_3$ group for the right-handed 
neutrinos like other fields,
the neutrino Yukawa coupling matrix $Y^{\nu}$
is restricted to the democratic form. Then the
mixing angles cannot be large.
The neutrino Yukawa matrix and the right-handed 
neutrino mass matrix are now parameterized as
\bea
Y^{\nu} &=& y^{\nu}_0(I + r J) ~+~ \Delta Y^{\nu}, \\
M_R &=& M_{R 0} (I + r' J) ~+~ \Delta M_R.
\eea
In should be noted that both of $Y^{\nu}$ and $M_R$ 
should be nearly diagonal in order to have large mixing angles .
Therefore the free parameters $r$ and $r'$
are constrained to be small, which is not explained by the flavor
symmetry.

Moreover this sort of fine-tuning turns out to 
be required much stronger in the GUT models.
In the SU(5) GUT, the quark and lepton fields 
in one generation
are combined into a ${\bf 10}+\bar{\bf 5}+{\bf 1}$
representation.
The superpotential of their Yukawa interactions is
given by
\be 
W = Y^u_{ij}~{\bf 10}_i {\bf 10}_j H({\bf 5})
+ Y^d_{ij}~{\bf 10}_i {\bf 5^*}_j H({\bf 5}^*)
+ Y^{\nu}_{ij}~{\bf 5^*}_i {\bf 1}_j H({\bf 5}).
\label{SU5}
\ee
We note that Yukawa couplings satisfy 
$Y^e = (Y^d)^T$ at the GUT scale.
Then $S_3$ symmetry for ${\bf 10}_i$ allows the Yukawa 
coupling matrix of $Y^u = y^u_0 (J + r'' I)$ in general.
However the mass hierarchy of up-sector quarks cannot be
obtained unless $r''$ is suppressed to $O(10^{-5})$.

Thus the approach of democratic mass matrix is an
attractive possibility phenomenologically, but the
flavor symmetries do not support origin of the assumed
Yukawa couplings.
In the next section, we consider realization of the
democratic Yukawa matrices in a dynamical fashion
without recourse to any flavor symmetries.
There it will be seen that we are not troubled with
the fine-tuning problems any more.

\section{Flavor democracy by strong gauge dynamics}
Our basic assumption is that the Yukawa couplings are not 
hierarchical nor aligned, but may be
even somewhat anarchy at the fundamental scale.
The purpose of this section is to show that 
the Yukawa couplings can be aligned to the rigid
democratic form quite minutely
at lower energy scale owing to strong gauge dynamics.
The realization of realistic matrices are discussed in
the later section.
Also we consider such a mechanism in GUT models,
since the problem of flavor symmetry is very
severe especially in GUT. 

\subsection{IR fixed point for a single Yukawa coupling}
Before going into the democratic Yukawa matrix,
let us consider the IR fixed point a la
Pendelton and Ross \cite{PR} in the cases with a single
Yukawa coupling.
In general, the RG equations for the gauge coupling $g$
and the Yukawa coupling $y$ are given at one-loop
level as
\bea
\mu \frac{d \ag }{d \mu} &=& -b\ag^2, \\
\mu \frac{d \ay }{d \mu} &=&
(a \ay - c \ag ) \ay,  
\eea
where we defined $\ag = g^2/ 8\pi^2$ 
and $\ay = |y|^2/8 \pi^2$.
The coefficients depend on the fields contents
and the Yukawa interaction, but note that 
$a$ and $c$ are positive constants. 
Then the ratio $x= \ay/ \ag$ satisfies the
RG equation,
\be
\mu \frac{d x }{d \mu} =
\left[ a x - (c - b) \right] \ag x.
\label{1x}
\ee
This beta function has a non-trivial
fixed point  at 
$x=x^* = (c-b)/a$, which
is IR attractive. 

The convergence behavior of the RG flows
around the fixed point may be seen by 
linear analysis.
The deviation from the fixed point,
$\Delta x = x - x^*$, is subject to
the equation given by
\be
\mu \frac{d \Delta x }{d \mu} =
(c - b) \ag \Delta x.
\ee
If the gauge coupling does not change
rapidly, then the flows of $x$ around the
IR fixed point may be evaluated as
\be
\Delta x(\mu) \sim
\left(\frac{\mu}{\Lambda}\right)^{(c-b)\ag}
\Delta x(\Lambda).
\label{convergence}
\ee
We see that the conditions for strong 
convergence as follows; 
(i) The  gauge coupling is large.
(ii) The coefficient $(c-b)$ is positive. 
Also larger $(c-b)$ makes convergence stronger.
The second condition means that the coefficient 
$b$ should not be very large, therefore, that 
asymptotically free gauge theories with rapidly running
gauge couplings are excluded.
\footnote{In asymptotically non-free gauge theories, 
it is easier to have strong convergence, since
the gauge couplings become large at high energy
and $b$ is negative\cite{nonfree}.
}

Such a fixed point appears in gauge theories beyond
four space-time dimensions as well\cite{powerlaw}.
In the dimensions of  $4+\delta$, 
the RG equations for the gauge coupling and
the Yukawa coupling at one-loop are given as
\bea
\mu \frac{d \ag }{d \mu} &=& \delta \ag - b \ag^2,  \\
\mu \frac{d \ay }{d \mu} &=&
\delta \ay + ( a \ay - c \ag ) \ay, 
\eea
where $\ag$ and $\ay$ denote dimensionless couplings.
Then the RG equation for $x= \ay/ \ag$ is found to be
identical to Eq.~(\ref{1x}). 
We should note some particular features in the 
extra-dimensional theories.
First the running law of these couplings is
not logarithmic but power with respect to the 
renormalization scale. Also
the gauge coupling becomes strong
at ultraviolet irrespectively of the sign of $b$.
With these reasons, convergence to the IR fixed
point is found to be very strong in general.

\subsection{Models with 3 flavors}
If we extend the Yukawa interaction to 
multi-flavors naively, then the IR fixed points
for the Yukawa matrices may exist but are proportional
to an identity matrix.
Thus the democratic Yukawa coupling matrix cannot
be obtained by renormalization effect in MSSM.
However it turns out to be possible once we admit
multi Higgs fields.

First let us consider only the up-type 
Yukawa couplings in the superpotential given 
by Eq.~(\ref{MSSM}) or by Eq.~(\ref{SU5}).
We introduce 9 elementary higgs superfields 
$H_{ij} (i, j = 1, 2, 3)$ with the same properties, 
and extend the Yukawa interactions as
\be
W = \sum_{i,j = 1,2,3} Y_{ij}  Q_i  u_j  H_{ij}.
\label{upYukawa}
\ee
In the SU(5) GUT, we may take $Q = u = {\bf 10}$
and H=H({\bf 5}).
In the supersymmetric theories the beta functions 
for the Yukawa couplings $Y_{ij}$ are written down 
in terms of the anomalous dimensions of
$Q_i$, $u_i$ and $H_{ij}$.
This is due to the so-called non-renormalization of
superpotentials.
Explicitly these anomalous dimensions may be written
down as
\bea
\gamma_{Q_i} &=& 
\left[ a_Q (\ay_{i1} + \ay_{i2} + \ay_{i3}) - c_Q \ag \right], \\
\gamma_{u_i} &=& 
\left[ a_u (\ay_{1i} + \ay_{2i} + \ay_{3i}) - c_u \ag \right], \\
\gamma_{H_{ij}} &=& 
\left[ 3 a_H \ay_{ij} - c_H \ag \right],
\label{anomalousdim}
\eea
where we defined  $\ay_{ij}= |Y_{ij}|^2/(8\pi^2)$ as well.
Note that the above interactions in Eq.~(\ref{upYukawa})
induce anomalous dimensions only in the diagonal elements,
which are given above.
Then the RG equation for $\ay_{ij}$ is simply given by
\be
\mu \frac{d \ay_{ij}}{d \mu} =
\left( \gamma_{Q_i} + \gamma_{u_j} + \gamma_{H_{ij}} \right)
\ay_{ij}.
\label{upYukawaRG}
\ee

It is easy to see existence of a non-trivial fixed point 
by using the beta function given by Eq.~(\ref{upYukawaRG}).
The RG equations for $x_{ij} = \ay_{ij}/\ag$ are
written down as
\be
\mu \frac{d x_{ij}}{d \mu} =
\left[ (b-c) + \left(
\hat{\gamma}_{Q_i} + \hat{\gamma}_{u_j} 
+ \hat{\gamma}_{H_{i,j}} \right)
\right] \ag x_{ij},
\label{9x}
\ee
where we have defined $c = c_Q + c_u + c_H$ and also
\bea
\hat{\gamma}_{Q_i} &=& 
a_Q ( x_{i1} + x_{i2} + x_{i3} ), \\
\hat{\gamma}_{u_i} &=& 
a_u ( x_{1j} + x_{2j} + x_{3j}), \\
\hat{\gamma}_{H_{ij}} &=& 
3 a_H x_{ij}.
\eea
Therefore the condition for the non-trivial
fixed point is that the combinations in the brace 
of the Eq.~(\ref{9x}) vanish 
for all sets of $(i, j)$.
The solution is unique and is found to be
\be
x_{ij}^* = x^* = \frac{c-b}{3a},
\label{9fp}
\ee
where $a= a_Q + a_u + a_H$.

The non-trivial but important matter
is whether this fixed point is
really IR attractive or not.
This may be seen by linear analysis of the RG
equations around the fixed point (\ref{9fp}).
The deviations from the fixed point
$ \Delta x_{ij}$ satisfy the following equation;
\be
\mu \frac{d \Delta x_{ij}}{d\mu} =
\ag x^* \left(
\begin{array}{ccccccccc}
a' & a_Q & a_Q & a_u & 0 & 0 & a_u & 0 & 0 \\
a_Q & a' & a_Q & 0 & a_u & 0 & 0 & a_u & 0 \\
a_Q & a_Q & a' & 0 & 0 & a_u & 0 & 0 & a_u \\
a_u & 0 & 0 & a' & a_Q & a_Q & a_u & 0 & 0 \\
0 & a_u & 0 & a_Q & a' & a_Q & 0 & a_u & 0 \\
0 & 0 & a_u & a_Q & a_Q & a' & 0 & 0 & a_u \\
a_u & 0 & 0 & a_u & 0 & 0 & a' & a_Q & a_Q \\
0 & a_u & 0 & 0 & a_u & 0 & a_Q & a' & a_Q \\
0 & 0 & a_u & 0 & 0 & a_u & a_Q & a_Q & a' 
\end{array}
\right)~\Delta x_{ij},
\label{linear9x}
\ee
where $a' = a_Q + a_u + 3 a_H$.
The eigenvalues of this matrix are found to be
$3 a_H$, $3 a_H$, $3 a_H$, $3 a_H$, $3(a_Q + a_H)$, 
$3(a_Q + a_H)$, $3(a_u + a_H)$, 
$3(a_u + a_H)$, $3 a$.
It should be noted that all of these are positive,
which ensures IR attractiveness of the fixed point.
Thus all Yukawa couplings may be aligned to the
same value by renormalization effect.
\footnote{To be precise, what is aligned at low energy is
not a Yukawa coupling but it's absolute value.
The complex phase is not controlled by the dynamics.
In this article, we treat the Yukawa couplings as if
real, and do not discuss the complex phases.}

It is possible also to reduce the number of Higgs upto 
three by assuming a $Z_3$ symmetry as follows.
We assign the $Z_3$-charge $q_i = 2\pi/3 i$ to 
$(Q_i, u_i, H_i) (i=1,2,3)$ respectively.
Then the $Z_3$ symmetry restricts the Yukawa interactions to
\be
W = \sum_{i,j = 1,2,3} Y_{ij} Q_i u_j H_{3-i-j},
\ee
where the indexes are defined modulo 3.
With these interactions the anomalous dimensions
are given to be diagonal.
Explicitly they are found to be
\bea
\gamma_{Q_i} &=& 
\left[ a_Q ( \ay_{i1} + \ay_{i2} + \ay_{i3}) - c_Q \ag \right], \\
\gamma_{u_i} &=& 
\left[ a_u ( \ay_{1i} + \ay_{2i} + \ay_{3i}) - c_u \ag \right], \\
\gamma_{H_i} &=& 
\left[ a_H ( \ay_{1(2-i)} + \ay_{2(1-i)} + \ay_{3(3-i)}) 
- c_H \ag \right].
\eea

The RG equations for  $x_{ij} = \ay_{ij}/\ag$ are given by
\be
\mu \frac{d x_{ij}}{d \mu} =
\left[ (b-c) + \left(
\hat{\gamma}_{Q_i} + \hat{\gamma}_{u_j} 
+ \hat{\gamma}_{H_{(3-i-j)}} \right)
\right] \ag x_{ij},
\ee
where $c= c_Q + c_u + c_H$ and $\hat{\gamma}= \gamma/\ag$ again.
It is immediately seen that couplings $x_{ij}^* = x^* = (c-b)/3a$
($a = a_Q +a_u +a_H$) give a set of fixed point solutions.
However it is found that this is not a unique solution.
The linear perturbation around the fixed point
$\Delta x_{ij}$ satisfies the differential equation,
\be
\mu \frac{d \Delta x_{ij}}{d\mu} =
\ag x^* \left(
\begin{array}{ccccccccc}
a & a_Q & a_Q & a_u & 0 & a_H & a_u & a_H & 0 \\
a_Q & a & a_Q & a_H & a_u & 0 & 0 & a_u & a_H \\
a_Q & a_Q & a & 0 & a_H & a_u & a_H & 0 & a_u \\
a_u & a_H & 0 & a & a_Q & a_Q & a_u & 0 & a_H \\
0 & a_u & a_H & a_Q & a & a_Q & a_H & a_u & 0 \\
a_H & 0 & a_u & a_Q & a_Q & a & 0 & a_H & a_u \\
a_u & 0 & a_H & a_u & a_H & 0 & a & a_Q & a_Q \\
a_H & a_u & 0 & 0 & a_u & a_H & a_Q & a & a_Q \\
0 & a_H & a_u & a_H & 0 & a_u & a_Q & a_Q & a 
\end{array}
\right)~\Delta x_{ij},
\label{Z3delta}
\ee
where $a= a_Q + a_u + a_H$. 
The eigenvalues of this matrix are found to be
$0$, $0$, $2a_Q$, $2a_Q$, $2a_u$, $2a_u$, $2a_H$, $2a_H$, 
$2a$.
Existence of two zero mode, namely constant modes, means
that the fixed point solution found above is not IR 
attractive. In other words, the Yukawa couplings obtained
at low energy are dependent on their initial values, which
are now supposed to be disordered.

However further flavor symmetry may relieve us of this
problem. Let us assume {\it e.g.} a discrete symmetry 
among matters,
\be
Q_i \rightarrow Q_{i+1}, ~~~
u_i \rightarrow u_{i-1}, ~~~
H_i \rightarrow H_i.
\ee
Then this symmetry imposes additional relations for the 
Yukawa couplings, which are given by
\be
x_{ij} = x_{(i+1)(j-1)} = x_{(i-1)(j+1)}.
\ee
Then it is found that the zero modes in Eq.~(\ref{Z3delta})
are just forbidden by this discrete symmetry.
Also linear perturbation shows us that the fixed point
is restricted to a unique one and is IR attractive.
Thus we have seen that the number of Higgs fields may be 
reduced by assuming some flavor structures in the fundamental
theory. However we do not consider this possibility more
in this article.

\subsection{Flavor democratic Higgs}
So far we have seen that the gauge dynamics may align the
Yukawa couplings to the same value at IR, once
many Higgs fields are introduced.
This does not lead us to a democratic mass matrix
immediately. Only one mode composed of many Higgs 
fields, which is identified with the Higgs field 
in the MSSM, should be massless and 
all others should be superheavy and
decoupled at low energy.
Moreover every fundamental Higgs field must contain
the massless mode by the same amount, otherwise
the Yukawa couplings are deformed from the democratic
ones.

In practice, we may construct the mass terms for the Higgs 
fields which ensure the above properties \cite{AK}.
We return to the 9 Higgs model given by Eq.~(\ref{upYukawa}).
In the case of SU(5) GUT, an example of the superpotential
for the Higgs fields may be given by
\bea
W & = & M
\sum_i (H({\bf 5})_{(i+1)j} - H({\bf 5})_{ij})
(H(\bar{\bf 5})_{(i+1)j} - H(\bar{\bf 5})_{ij}) \nn \\
& & + M
\sum_j (H({\bf 5})_{i(j+1)} - H({\bf 5})_{ij})
(H(\bar{\bf 5})_{i(j+1)} - H(\bar{\bf 5})_{ij}).
\eea
Then the mass terms for the Higgs scalar fields,
\be
M^2 \sum_i |H({\bf 5})_{(i+1)j} - H({\bf 5})_{ij}|^2
+ 
M^2 \sum_j |H({\bf 5})_{i(j+1)} - H({\bf 5})_{ij}|^2
+ \cdots,
\ee
give rise to just one massless mode.
Explicitly the massless mode $H$ is given by
\be
H = \frac{1}{3} \sum_{i,j} H_{ij}.
\ee
Presence of the massless mode is ensured by the shift symmetry
with transformations
\footnote{This structure is common to the deconstruction 
models\cite{deconstruction},
where the massless mode is generated as a Nambu-Goldstone boson.},
\be
H({\bf 5})_{i,j}\rightarrow C,~~~
H(\bar{\bf 5})_{i,j}\rightarrow \bar{C},
\ee
where $C$ and $\bar{C}$ are constants.
Other modes acquire mass of order $M$, which
gives decoupling scale.
In the later section, we shall take $M$ to be 
the GUT scale.

This massless modes composed of $H({\bf 5})$ and 
$H(\bar{\bf 5})$ are  identified with Higgs fields 
$H^u$ and $H^d$ in the MSSM respectively. 
It is important to note that the Higgs field
$H_{ij}$ in the both sector contains the zero mode $H$
with same factor
\be
H_{ij} = \frac{1}{3} H + \cdots,
\ee
for all $(i, j)$.
Therefore the Yukawa matrix appearing in the MSSM
is given to be democratic one, once the Yukawa couplings
have been aligned enough by the strong gauge dynamics.

It may be said that the basic policy in our 
consideration is flavor democracy.
We do not assume any hierarchical structures in
the initial Yukawa couplings. 
In addition the superpotential for Higgs field
does not have flavor difference.
Therefore, there is no essential difference
among flavors apriori. 

\section{A model based on the $SU(5) \times SU(5)$ GUT}
Next we consider the mass hierarchy between the first two 
generations and the third generation.
The small masses for the first and the second generations are
generated by slight deviations from the rigid
democratic mass matrix.
From our standpoint of flavor democracy, these deviations
should be traces of disorder in the initial couplings.
Mass ratio between up and top quarks is more than 
$1 : 10^5$.
On the other hand, ratio between the GUT scale and the Planck
scale, which we suppose to be the fundamental scale, is only
about $1 : 10^3$.
Therefore we need very strong gauge interaction in order
to achieve such a hierarchy. We may estimate roughly 
strength of the gauge coupling by assuming that it is not
running. It is seen from Eq.~(\ref{convergence}) that 
the gauge coupling must satisfies
\be
\frac{g^2}{(4\pi)^2} > \frac{5}{6(c-b)}.
\ee
This condition implies that the gauge coupling need to be 
non-perturbatively large.

Naively the gauge coupling constant of the SU(5) GUT
is supposed to be rather weak, since
unification of the three gauge couplings in the
MSSM occurs at rather weak coupling regime.
Therefore it seems unlikely to make the gauge
coupling of GUT so strong.
Actually this is not the case.
The gauge coupling unification is not destroyed in the
presence of any extra heavy fields belonging to
SU(5) multiplets.
However corrections by the extra matters enlarge
the unified gauge coupling.
Thus one way to realize strong unification is to assume
a suitable number of extra heavy fields in the MSSM \cite{AK}.

Here we consider another scenario which may be more
attractive
\footnote{One may expect also that power law running 
of gauge coupling in extra dimensions offers us models
with strong convergence as well. 
However we do not pursuit for this direction 
due to a problem discussed in section 7.}. 
Suppose that the gauge group is given by a
product of two $SU(5)$ groups and is broken spontaneously
to their diagonal subgroup at the unification scale.
We represent the gauge couplings for the two groups
$SU(5)'$ and $SU(5)''$ by $g'$ and $g''$ respectively.
Here we assume that all of the matter fields as well as
the Higgs fields are charged under
$SU(5)'$ but singlet under $SU(5)''$ and, moreover, that
that $g'$ is non-perturbatively strong but $g''$ is
weak. 
Then RG for the Yukawa couplings is driven by the 
$SU(5)'$ gauge interaction, and may be attracted to
their IR fixed point very rapidly.
Also the gauge coupling $g$ for the diagonal subgroup,
which is given by $1/g^2 = 1/g'^2 + 1/g''^2$, is given
weak. 

The superpotential to be considered here is the same 
as that given in Eq.~(\ref{SU5}) except for the 
Higgs fields extended to multi ones;
\be 
W = Y^u_{ij}~{\bf 10}_i {\bf 10}_j H({\bf 5})_{ij}
+ Y^d_{ij}~{\bf 10}_i \bar{\bf 5}_j H(\bar{\bf 5})_{ij}
+ Y^{\nu}_{ij}~\bar{\bf 5}_i {\bf 1}_j H({\bf 5})_{ij}.
\label{udSU5}
\ee
To be explicit, the anomalous dimensions for the
matter fields and the Higgs fields are given at 
one-loop level as follows;
\bea
\gamma_{{\bf 10}_i} &=&
- \frac{36}{5} \ag' + 3 \sum_{j=1}^3 \ay^u_{ij} 
+ 2 \sum_{j=1}^3 \ay^d_{ij},  \\
\gamma_{\bar{\bf 5}_i} &=&
- \frac{24}{5} \ag' + 4 \sum_{j=1}^3 \ay^d_{ji} 
+  \sum_{j=1}^3 \ay^{\nu}_{ij}, \\
\gamma_{{\bf 1}_i} &=& 
5 \sum_{j=1}^3 \ay^{\nu}_{ji},  \\
\gamma_{H({\bf 5})_{ij}} &=&
- \frac{24}{5} \ag' + 6  \ay^u_{ij} 
+ \ay^{\nu}_{ij},  \\
\gamma_{H(\bar{\bf 5})_{ij}} &=&
- \frac{24}{5} \ag' + 4  \ay^d_{ij}, 
\eea
where $\ag = g'^2/8\pi^2$ and
$\ay^a_{ij} = |Y^a_{ij}|^2/8\pi^2$ for $a= u, d, \nu$.
The RG equations for the Yukawa couplings 
$Y^u_{ij}, Y^d_{ij}$ and $Y^{\nu}_{ij}$ are written down
in terms of these anomalous dimensions as
\bea
\mu \frac{d \ay^u_{ij}}{d \mu} &=&
\left( \gamma_{{\bf 10}_i} + \gamma_{{\bf 10}_j} 
+ \gamma_{H({\bf 5})_{ij}} \right) \ay^u_{ij}, \\
\mu \frac{d \ay^d_{ij}}{d \mu} &=&
\left( \gamma_{{\bf 10}_i} + \gamma_{\bar{\bf 5}_j} 
+ \gamma_{H(\bar{\bf 5})_{ij}} \right) \ay^d_{ij}, \\
\mu \frac{d \ay^{\nu}_{ij}}{d \mu} &=&
\left( \gamma_{\bar{\bf 5}_i} + \gamma_{{\bf 1}_j} 
+ \gamma_{H({\bf 5})_{ij}} \right) \ay^{\nu}_{ij}.
\label{SU5RG}
\eea
In the next section we consider a dynamical mechanism 
compatible with large mixing angles in the lepton
sector. There all of the neutrino Yukawa couplings 
$Y^{\nu}_{ij}$ are made decrease to very small values.
With this reason, we shall consider RG behavior of other
Yukawa couplings by neglecting the neutrino Yukawa 
couplings in this section.

Now two kinds of Yukawa couplings $\ay^u$ and $\ay^d$
are coupled with each other in the RG equations.
Regardless of this complication, it is found that a 
non-trivial fixed point exists and is IR attractive.
The coupling ratios $x^u_{ij}=\ay^u_{ij}/\ag$ and 
$x^d_{ij}=\ay^d_{ij}/\ag$ satisfy the following  
equations,
\bea
\mu \frac{d x^u_{ij}}{\mu} &=&
\left(
-\frac{96}{5} + b 
+ \sum_{k=1}^3 \left(3 x^u_{ik} + 3 x^u_{jk}
+ 2 x^d_{ik} + 3 x^d_{jk} \right)
+6 x^u_{ij}
\right)\ag' x^u_{ij}, \label{ux} \\
\mu \frac{d x^d_{ij}}{\mu} &=&
\left(
-\frac{84}{5} + b 
+ \sum_{k=1}^3 \left(3 x^u_{ik} + 2 x^d_{ik} 
+ 4 x^d_{kj}\right) + 4 x^d_{ij}
\right)\ag' x^d_{ij}.
\label{dx}
\eea
Then it is straightforward to find the non-trivial 
fixed point solution, which turns out to be
\bea
x^u_{ij} &=& x^{u*} = \frac{552-25 b}{1050} \sim 0.53 - 0.02 b, 
\label{ufp} \\
x^d_{ij} &=& x^{d*} = \frac{384-25 b}{700} \sim 0.55 - 0.04 b.
\label{dfp}
\eea 
Thus the Yukawa couplings are fixed at the GUT scale, once
$b$ is given. In any case, the Yukawa couplings become
non-perturbatively large accompanied with the gauge coupling.
It is a rather tedious problem to verify the IR attractive 
nature of this fixed point by linear perturbation.
However it is quite obvious to see it by solving the differential
equations given by (\ref{ux}) and (\ref{dx}) numerically.
In Fig.~1, the flow lines for 
$x^u_{ij}$ and $x^d_{ij}$ are shown in the case of 
$b=0, \ag'=1.0$. 
Their initial conditions are chosen at random just for
demonstration.
It is seen that both couplings converge their fixed point
values very rapidly.

\newpage
\begin{figure}[thb]
\begin{center}
\epsfxsize=0.6\textwidth
\leavevmode
\epsffile{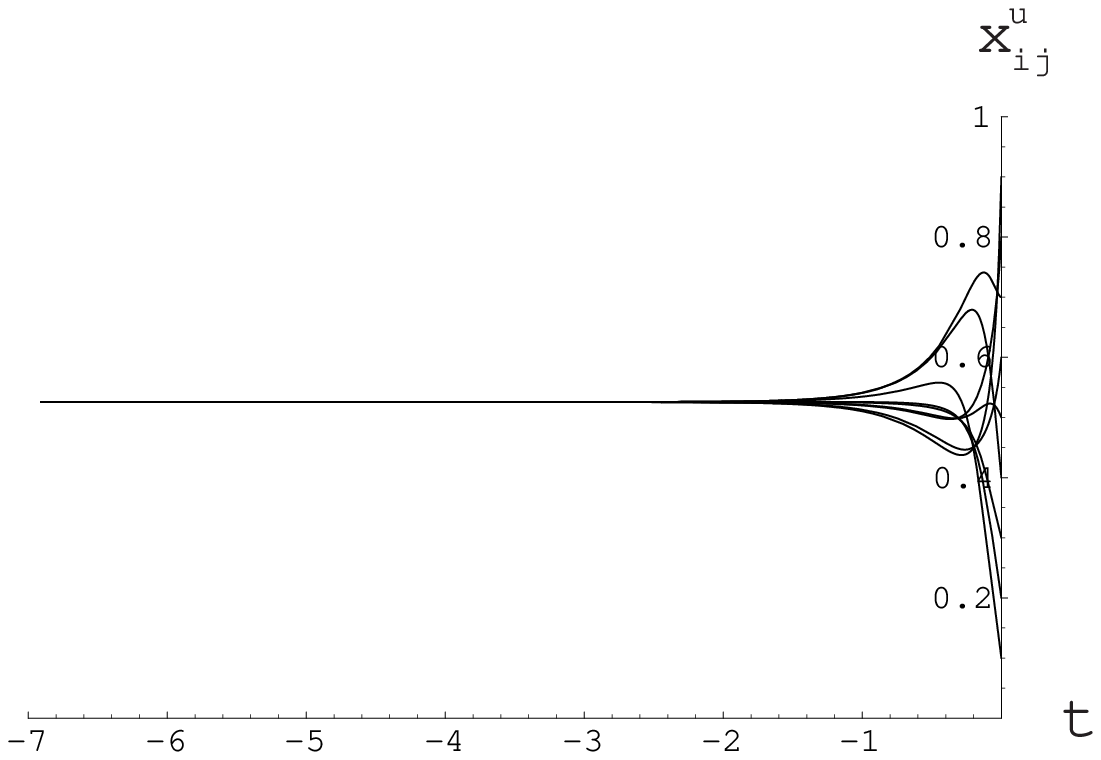}
\epsfxsize=0.6\textwidth
\leavevmode
\epsffile{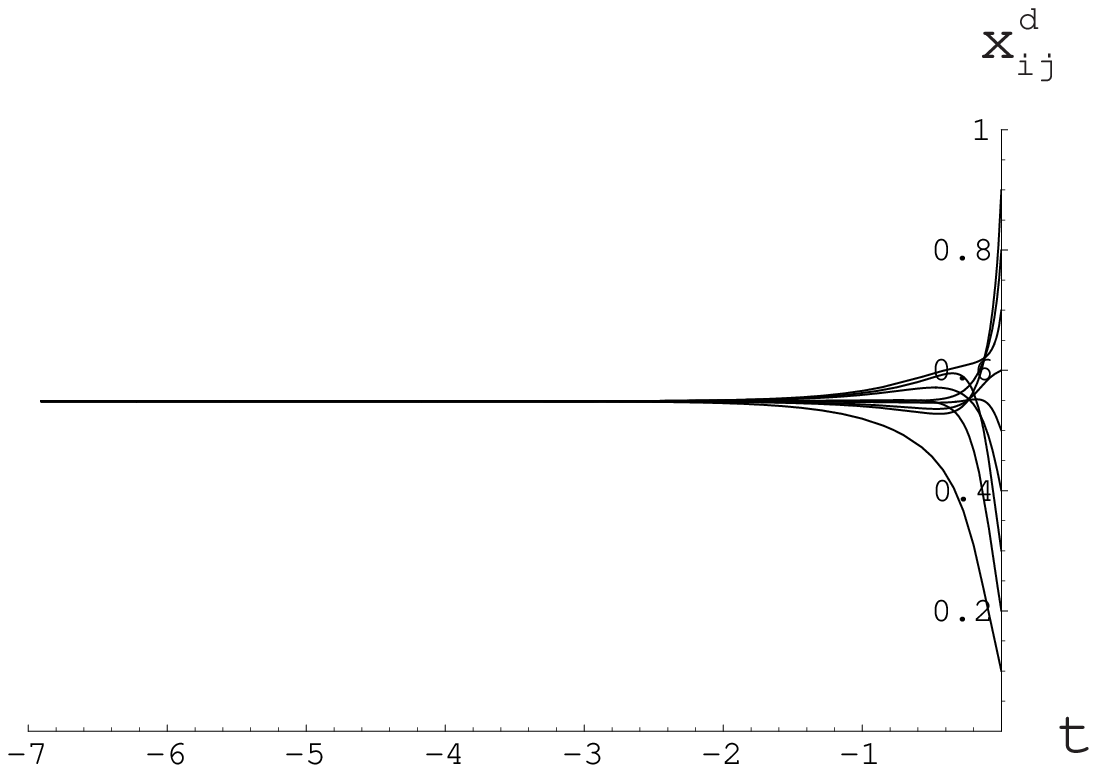}
\label{fig1}
\caption{RG running of $x^u_{ij}$ and $x^d_{ij}$
for a sample of the initial couplings chosen at random.
We assumed also $b=0, \ag'=1.0$.
The parameter $t$ represent the
renormalization scale $\mu$ by using
$\ln (\mu/M_{\rm pl})$, where $M_{\rm pl}$ is the
Plank scale.
}
\end{center}
\end{figure}

People have studied GUT models based on a product group 
for the purpose of avoiding the doublet-triplet splitting 
problem \cite{DT,witten}.
Recently models based on $G=SU(5)' \times SU(5)''$ also
have been proposed in this context \cite{witten},
where gauge charge assignment for Higgs fields is
different from the above setting. It is assumed that
$H(\bar{\bf 5})$ carries $({\bf 1}, \bar{\bf 5})$
charges for the product group $SU(5)' \times SU(5)''$,
while $H(\bar{\bf 5})$ carries $({\bf 5}, {\bf 1})$
as before.
It is noted that the down-type Yukawa interactions in the
superpotential given by Eq.~(\ref{udSU5}) are not
invariant any more.
In order to generate the down-type Yukawa interactions,
we introduce another field $\Sigma$ belonging to
$(\bar{\bf 5}, {\bf 5})$.
We suppose also that the spontaneous symmetry breaking
$SU(5)' \times SU(5)'' \rightarrow SU(5)_{\rm diag}$
is induced by a vacuum expectation value of $\Sigma$.
Then an invariant, but non-renormalizable, term
\be
\frac{Y^d_{ij}}{\Lambda}~
{\bf 10}_i \bar{\bf 5}_j H(\bar{\bf 5})_{ij} \Sigma
\ee
may generate the down-type Yukawa couplings effectively 
after symmetry breaking.
It seems rather difficult to discuss renormalization involved
with non-renormalizable operators like this.
However we may also consider the corrections at the broken
vacuum, where these operators plays a role of
the Yukawa interactions instead.
Therefore it is expected that the above analysis for the 
IR fixed point is equally applied to this model.

As a general property of the present mechanism, the Yukawa
couplings for the third generations are given to be 
rather large at the GUT scale.
The masses for top, bottom and tau are given in terms of
the Yukawa couplings obtained after diagonalization.
It is easy to find these Yukawa couplings renormalized at 
$O(100)$GeV  by solving RG equations for the MSSM.
The neutrino Yukawa couplings may well be neglected
in this analysis.
Then the low energy couplings are determined with respect to 
the strong gauge coupling $g'$ and the model parameter $b$.
However it is found that the resultant Yukawa couplings 
are almost insensitive to these parameters.
This is because the initial Yukawa couplings are fairly large
and such flows converge to the so-called quasi-fixed point
\cite{quasifp}.
To be explicit, the low energy couplings are found to be
$Y_t=1.00$, $Y_b=0.94$ and $y_{\tau}=0.62$ in the case of
$\ag'=g'^2/8\pi^2 = 1.0$ and $b=0$.
It is  seen first that $\tan \beta$ should be large as
$O(50)$, to explain the mass ratio $m_t/m_b$.
since $Y_t$ and $Y_b$ appear to be almost the same.
Therefore mass of top quark is predicted to be the same as
the vacuum expectation value of
the neutral Higgs $174$GeV,
which shows a very good agreement with observation. 
We may also predict mass ratio of bottom and tau,
$m_b/m_{\tau}$.
The above couplings bring us to 
$m_b/m_{\tau} = 1.52$,
which is slightly smaller than the expected value.\footnote{
In the case with large $\tan \beta$, we have to take 
into account large SUSY threshold corrections on the 
bottom mass\cite{hall}.}

\section{Neutrino sector}
In the previous section we have seen that the Yukawa couplings
are aligned to the democratic forms very well.
The deviations from the rigid democratic form are
responsible for masses of the first two generations and
mixing angles among quarks.
However the mixing matrix becomes necessarily close to
an identity matrix for any small deviations.
If we take the neutrino Yukawa couplings, ignored in the 
previous section, into account, then these couplings are found to 
be attracted  to an IR fixed point by strong dynamics as well as
other Yukawa couplings.
Obviously this is incompatible with the observed large mixing 
angles for leptons.

In this respect, it is phenomenologically favorable
for the neutrino Yukawa matrix  to be nearly 
diagonal at low energy.
Suppose that the neutrino Yukawa couplings do not have 
a non-trivial fixed point with some reasons. 
Then these couplings are not aligned to the democratic
form and become dependent on their initial couplings
at the Planck scale. Also we may find a sizable 
parameter region for the initial couplings consistent 
with the observed neutrino mass and mixings.

In practice such a situation turns out to be realized
with a simple assumption, 
since the right-handed neutrino is a
singlet of the $SU(5)$.
In general, GUT models may contain a pair of superheavy 
vector-like fields and so on.
Then the right-handed neutrino is allowed to have 
additional Yukawa interactions with them
\footnote{R-parities for the extra fields may be 
assigned properly.}.
These Yukawa couplings are driven to be very large 
like others by the strong gauge interactions.
Consequently anomalous dimension of the 
right-handed neutrino is enhanced significantly.
We note that the IR fixed point stands upon balance
of negative contribution by gauge interactions and
positive contribution by Yukawa interactions.
Presence of extra Yukawa couplings may destroy
this balance and take the IR fixed point away
from the neutrino Yukawa couplings in the end.
Explicitly the coefficient $c$ in the fixed point 
equation (\ref{9x}) is made negative by the
additional radiative corrections.
Thus the remarkable difference in the mixing matrices of
quark and lepton sectors may attribute to the
dynamics of the right-handed neutrino.

We shall demonstrate the above mechanism by examining
the RG equations only for the neutrino Yukawa couplings.
As a toy model with a single flavor, 
let us consider the superpotential given by
\be
W = Y^{\nu} ~\bar{\bf 5} ~{\bf 1} ~H({\bf 5})
+ \kappa  {\bf 1} ~\Phi ~\bar{\Phi},
\ee
where the extra matter $\Phi$ and $\bar{\Phi}$ belong
to $SU(5)$ representations ${\bf R}$ and $\bar{\bf R}$
respectively.
The RG equations for $\ay=|Y^{\nu}|^2/8\pi^2$,
and $\ak=|\kappa|^2/8\pi^2$  are found to be 
\bea
\mu \frac{d \ay}{d \mu}&=&
\left[7\ay + R \ak - 4 C_2({\bf 5}) \ag \right] \ay, \\
\mu \frac{d \ak}{d \mu}&=&
\left[5\ay + (R+2) \ak - 4 C_2({\bf R}) \ag \right] \ak,
\eea
where $C_2$ denotes the Casimir index of each representation.
Also these are rewritten into equations in terms of
$x_y = \ay/\ag$ and $x_{\kappa} = \ak/\ag$ as
\bea
\mu \frac{d x_y}{d \mu}&=&
\left[7 x_y + R x_{\kappa} - 4 C_2({\bf 5}) + b \right] \ag x_y, \\
\mu \frac{d x_{\kappa}}{d \mu}&=&
\left[5 x_y + (R+2) x_{\kappa} - 4 C_2({\bf R}) + b \right] 
\ag x_{\kappa}.
\eea
There are four fixed points in the coupling space of
$(x_y, x_{\kappa})$, three of which are immediately seen
from the above equations as
\bea
(x_y, x_{\kappa}) &=& (0,0),  \label{fpA} \\
&=& \left( \frac{4 C_2({\bf 5}) - b}{7} ,0 \right), \label{fpB}\\
&=& \left( 0 ,  \frac{4 C_2({\bf R}) - b}{R+2} \right). \label{fpC}
\eea
The first two fixed points are what we have been
discussing so far and the third one is a new entry. 
It is noted that the fourth one appears in the
region of negative $\ay$ for a large representation
${\bf R}$
\footnote{
This point may not be unphysical with taking complex 
phases of the Yukawa coupling into considerations.
However this fixed point is not IR attractive anyway. }.
Here we pay attention to such a case.

The RG flow diagram for $(x_y, x_{\kappa})$ is shown 
in Fig.~2 in the case of $b=0$ and ${\bf R}= {\rm \bf adj}$ 
as an example.
The marked points A, B and C stands for the fixed points 
given by (\ref{fpA}), (\ref{fpB}) and (\ref{fpC})
respectively.
It is seen that the fixed point B is now unstable 
towards direction of the new coupling, 
while C turns out to be IR attractive instead.
Thus the neutrino Yukawa coupling is found to 
decrease at low energy in the presence of $\kappa$.

\begin{figure}[thb]
\begin{center}
\epsfxsize=0.6\textwidth
\leavevmode
\epsffile{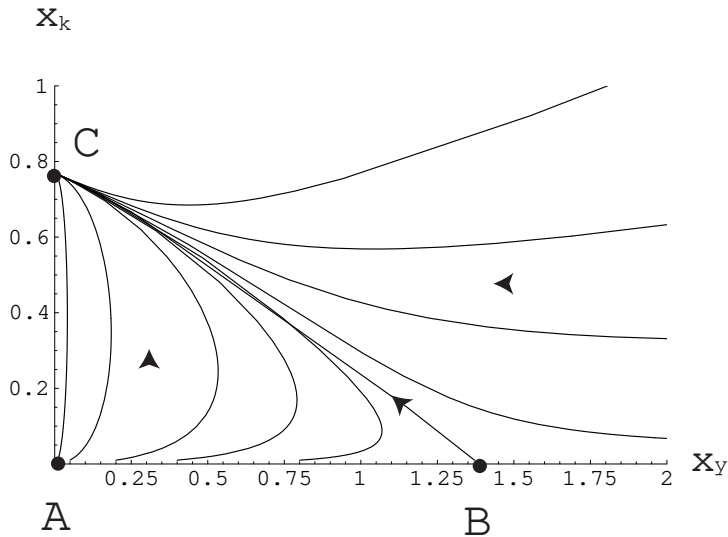}
\caption{RG flows for $(x_y, x_{\kappa})$ in the case of 
$b=0$ and ${\bf R}= {\rm \bf adj}$. The arrows show
the direction to lower energy scale. The circles at
A, B and C stand for the fixed points given by
(\ref{fpA}), (\ref{fpA}) and (\ref{fpC}) respectively.}
\end{center}
\label{fig3}
\end{figure}

In Fig.~3 evolution of neutrino Yukawa coupling with
respect to the scale is shown.
Each line corresponds with a RG flow presented in Fig.~2.
It is found that the Yukawa couplings decrease monotonically
at low energy and are not aligned.
It is expected that the neutrino Yukawa couplings for
multi-flavors also show the same aspect of these flows,
even though other Yukawa couplings $Y^u$ and $Y^d$
are aligned to the fixed point.
We now add the following terms,
\be
Y^{\nu}_{ij} \bar{\bf 5}_i {\bf 1}_j  H({\bf 5})_{ij}
+ \kappa_i  {\bf 1}_i \Phi ~\bar{\Phi}
+ M_{R ij} {\bf 1}_i  {\bf 1}_j,
\ee
to the superpotential given by Eq.~(\ref{udSU5}).
It is easily found that ratio of the neutrino Yukawa 
couplings follows the RG equation given by
\bea
\mu \frac{d}{d \mu} 
\ln \left(
\frac{\ay^{\nu}_{ik}}{\ay^{\nu}_{jk}}
\right)
&=&\left(
\gamma_{{\bf 5}^*_i} + \gamma_{{\bf 1}_k} + 
\gamma_{H({\bf 5})_{ik}}
\right)
- \left(
\gamma_{{\bf 5}^*_j} + \gamma_{{\bf 1}_k} + 
\gamma_{H({\bf 5})_{jk}}
\right) \nn \\
&=& a_{\bar{\bf 5}} 
\left( \sum_k \ay^{\nu}_{ik} - \ay^{\nu}_{jk} \right)
+ 3 a_H \left( \ay^{\nu}_{ik} - \ay^{\nu}_{jk} \right),
\label{nonalign}
\eea
where the Yukawa couplings other than $\ay^{\nu}$ are
assumed to be their fixed point values.
Since the neutrino Yukawa couplings are suppressed
as seen above, the right-hand side of (\ref{nonalign})
becomes very small quickly. 
This shows that ratio of the neutrino couplings is
almost unchanged  at low energy region.

\begin{figure}[thb]
\begin{center}
\epsfxsize=0.6\textwidth
\leavevmode
\epsffile{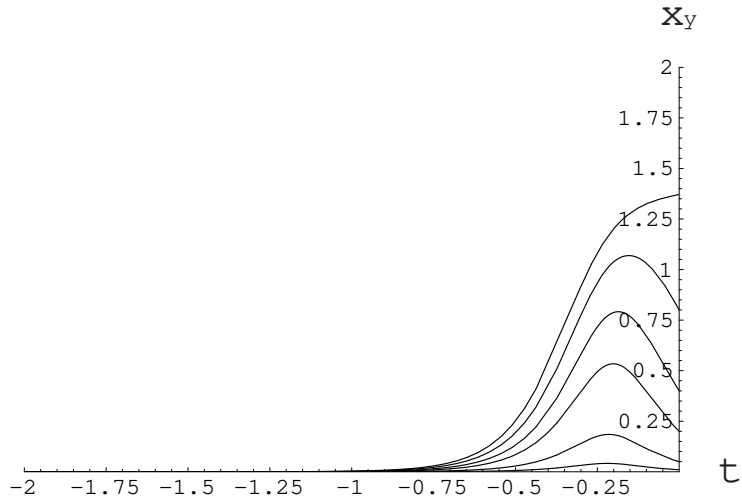}
\caption{Running of $x_y$ corresponding with
various flow lines shown in Fig.~2 with respect to
the scale $\mu$ ($t = \ln(\mu/M_{\rm pl})$).}
\end{center}
\label{fig4}
\end{figure}

In the Frogatt-Nielsen mechanism \cite{FN}, the neutrino 
Yukawa couplings are constrained upto O(1) coefficient,
and, therefore, the masses and mixing angles 
for neutrinos are not predicted.
However we may find a sizable probability for the couplings
distributed in anarchy to be compatible with the observed
masses and mixing angles \cite{anarchy}.

Now if three of nine initial neutrino Yukawa couplings happen to be
larger than others to some extent, then they may be dominant
in the low energy matrix also.
Since there is no specific flavor basis, we may regard
these couplings as the diagonal elements.
Thus it is not so special that the neutrino Yukawa coupling
matrix appears close to diagonal, and therefore the large
lepton mixing angles are described well.
We leave explicit survey of such a parameter space for
future works.

In our mechanism the neutrino Yukawa couplings are
largely suppressed  as seen in Fig.~3.
However this does not imply that the see-saw mechanism offers
us very tiny neutrino masses.
Note that this suppression occurs because of
enhancement of anomalous dimensions of the right-handed
neutrinos.
Then their Majorana masses $M_R$ are also suppressed by
the anomalous dimensions.
Consequently it is found that the neutrino masses 
themselves are not suppressed through this mechanism.
It is noted that 
the masses for the right-handed neutrinos
are expected to appear at the intermediate scale
in our scenario, even though the bare parameter for $M_R$ 
is set to be the GUT scale.

\section{Setup for the realistic mass matrices}
How can we describe mass differences among the first
and the second generations at all?
In the phenomenological approach, the two flavor symmetry
breaking parameters $\epsilon$ and $\delta$ are
taken to be hierarchical as given in Eq.~(\ref{breakings}).
Besides it is necessary for these breaking parameters 
to be diagonal elements, which is not explained from
the standpoint of flavor symmetry either.

Now the problem to be considered is how to realize such
a specific form of deviations from the democratic couplings
in our dynamical framework.
The first possibility is to generate such deviations away
from the fixed point by assuming proper initial
couplings.
We consider this by examining the RG equations given 
by Eq.~(\ref{9x}) again.
Suppose, {\it e.g.} that one of the Yukawa couplings,
say $Y_{33}$, happens to be fairly 
smaller than the others at the Planck scale.
Then $x_{ij}$ for $i, j = 1, 2$ converge into
the fixed point value rapidly, since the beta
functions for these couplings do not contain
$x_{33}$.
The couplings $x_{i3}$ and $x_{3i}$ for $i = 1, 2$
may be affected by $x_{33}$.
So we may examine the RG flows with neglecting
differences among $x_{ij}$ for $i, j = 1, 2$
in the first step of approximation.
In this situation the deviations from the fixed point 
$\Delta x_{ij}= x_{ij} - x^* $ are found to satisfy 
\bea
\mu \frac{d}{d \mu}\left[ \Delta x_{13} - \Delta x_{23} \right]
&=& \ag x^* (a' - a_u) \left[ \Delta x_{13} - \Delta x_{23} \right] 
\label{diffi3} \\
\mu \frac{d}{d \mu}\left[ \Delta x_{31} - \Delta x_{32} \right]
&=& \ag x^* (a' - a_Q) \left[ \Delta x_{31} - \Delta x_{32} \right].
\label{diff3i}
\eea
These equations tell us that these differences shrink rapidly 
with scaling down and eventually the couplings satisfy the relations
\be
x_{13}=x_{23},~~~x_{31}=x_{32}.
\ee
Moreover differences among $x_{ij}$ for $i, j = 1, 2$,
which is ignored in the above argument, are found to
satisfy
\bea
\mu \frac{d}{d \mu}\left[ \Delta x_{11} - \Delta x_{21} \right]
&=& 
\mu \frac{d}{d \mu}\left[ \Delta x_{12} - \Delta x_{22} \right]
=
\ag x^* a_Q \left[ \Delta x_{13} - \Delta x_{23} \right],  
\label{diff12} \\
\mu \frac{d}{d \mu}\left[ \Delta x_{11} - \Delta x_{12} \right]
&=& 
\mu \frac{d}{d \mu}\left[ \Delta x_{21} - \Delta x_{22} \right]
=
\ag x^* a_u \left[ \Delta x_{31} - \Delta x_{32} \right].
\label{diff21}
\eea
Therefore it is seen that these differences are not affected 
from large deviation of $x_{33}$ and are kept small.
In Fig.~4 a sample of RG flows for the couplings $x_{ij}$
obtained by numerical analysis are demonstrated.  
The bold line shows flow of $x_{33}$, whose initial
coupling is assumed to be relatively smaller than others.
Note that Yukawa coupling $Y_{33}$ itself is not so 
apart from others.
The dashed lines represent flow of $x_{i3}$ and $x_{3i}$
for $i = 1, 2$, while the dotted lines represent flow
of $x_{ij}$ for $i, j = 1, 2 $.
Indeed behavior of these RG flows supports the above
argument.

\begin{figure}[thb]
\begin{center}
\epsfxsize=0.6\textwidth
\leavevmode
\epsffile{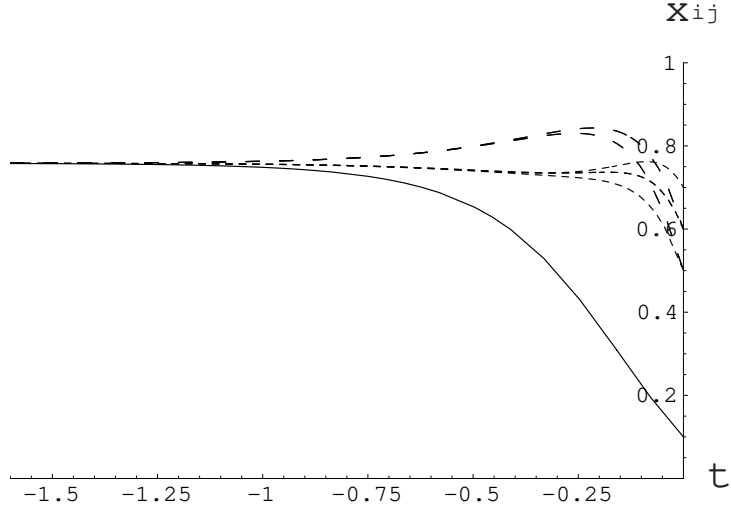}
\caption{Convergence behavior of $x_{ij}$ to the IR fixed
point in the case of a proper initial couplings. 
The parameter $t$ denotes $\ln (\mu/M_{\rm pl})$.
The bold line shows flow of $x_{33}$, whose initial
coupling is assumed to be rather smaller than others.
The dashed lines represent flow of $x_{i3}$ and $x_{3i}$
for $i = 1, 2$, while the dotted lines represent flow
of $x_{ij}$ for $i, j = 1, 2 $.
}
\end{center}
\label{fig5}
\end{figure}

Similar forms of Yukawa coupling matrices may be obtained 
by introducing a weak interaction breaking the flavor democracy.
For example, we assume only one of the Higgs fields, say
$H_{33}$, has an extra superpotential
$\lambda H_{33} \Phi_1 \Phi_2$ with very small coupling 
$\lambda$.
We consider the linear perturbation around the
fixed point.
The extra Yukawa interaction affects only on
the beta function of $\Delta x_{33}$.
To be explicit, the RG equations given in Eq.~(\ref{linear9x})
are modified to 
\be
\mu \frac{d \Delta x_{ij}}{d \mu} =
\ag x^* \left[ ({\cal M} \Delta x)_{ij} 
+ \delta_{\lambda} \delta_{i3} \delta_{j3}
\right],
\ee
where ${\cal M}$ denotes the matrix given in Eq.~(\ref{linear9x})
and $\delta_{\lambda} \ag x^* \sim |\lambda|^2/8\pi^2$.
When differences among the initial Yukawa couplings happen to be
rather small already, the same equations given by 
Eqs.~(\ref{diffi3}), (\ref{diff3i}), (\ref{diff12})
and (\ref{diff21}) hold as well.

In the end, the Yukawa couplings turn out to be 
represented by the form of
\be 
Y  =y_0
\left(
\begin{array}{ccc}
1 + O(\epsilon) & 1+ O(\epsilon) & 1+ \delta' \\
1 + O(\epsilon) & 1+ O(\epsilon) & 1 +\delta' \\
1 +\delta'' & 1 +\delta'' & 1 +\delta
\end{array}
\right),
\ee
at low energy scale.
Here $\epsilon$ stands for the incompleteness of
strong convergence to the fixed point,
which can be as small as $10^{-5}$.
Other parameters representing deviations from the
fixed point $\delta$, $\delta'$ and $\delta''$
are supposed to be of the same order.
Though these values are dependent on the
initial Yukawa couplings, we may well assume
$\epsilon \ll \delta \ll 1$.
After transforming this matrix $Y$ by using
the diagonalizing matrix given by Eq.~(\ref{A-matrix}),
we obtain
\be 
A^T Y A =y_0
\left(
\begin{array}{ccc}
O(\epsilon) & O(\epsilon) & O(\delta) \\
O(\epsilon) & O(\delta) & O(\delta) \\
O(\delta) & O(\delta) & 3 + O(\delta) \\
\end{array}
\right).
\ee
This is very similar to the Fritzsch type matrix and,
therefore, offers us a phenomenologically 
favorable pattern for realistic masses and mixing angles.
The ratio mass eigenvalues is given by 
$O(\epsilon) : O(\delta) : O(1)$.
We do not study which initial couplings really give
the viable mass matrices explicitly, since there are
large model dependence.

\section{Conclusions and discussions}
The approach of democratic mass matrix seems to be attractive
enough phenomenologically.
The notion of flavor democracy offers us an interesting
viewpoint distinct from other approach.
Above all it is noted that the lepton mixing angles,
bi-large  mixings and small $U_{e3}$, are
well described by assuming a democratic mass matrix
for the charged leptons.

However it is also true that these phenomenologically
favorable matrices are not explained by flavor symmetries.
In this paper we considered the models in which the democratic
type of Yukawa couplings are realized as an IR attractive fixed 
point\cite{AK}.
For this purpose we introduced multi Higgs fields, from which
only one massless mode is allowed by a special type of 
mass terms.
The large mass hierarchies are achieved easily by using
strong gauge interaction for quarks and leptons as well as
the Higgs fields.
To be explicit, we considered an $SU(5) \times SU(5)$ GUT
model, in which one of the gauge couplings is 
non-perturbatively strong.

Indeed the setup of the Higgs fields may be somewhat artificial.
However we found some benefits in this scenario.
The democratic Yukawa couplings are realized even in GUT 
models without any adhoc assumptions.
Hence small mixing angles as well as large mass hierarchies
in the quark sector are naturally obtained.
Especially the large lepton mixing angles
can be explained through  extra interactions of the
right-handed neutrinos.
It is seen that simple setting of the initial couplings
may lead to low energy Yukawa couplings of the Fritzsch type.
Also this kind of scenarios predict top quark mass in the
good agreement with experiment and large $\tan \beta$.

In this article we considered only supersymmetric models.
However the PR fixed point exists also in non-supersymmetric
theories as well. Therefore the democratic mass matrices may
be realized also in some non-supersymmetric models, though
the beta functions for Yukawa couplings are complicated.

In section 3 we mentioned that the PR fixed point appears 
also in extra dimensions and shows rather strong convergence
in general. 
However naive extension with gauge and Higgs in the
bulk of the extra dimensions does not work.
This is because anomalous dimensions of the Higgs fields
vanish due to effective N=2 supersymmetry,
and ,therefore, the democratic type of fixed point
is not realized.
This would not deny any possibility for extra 
dimensional models, and we leave it for future study.

Lastly some comments on soft supersymmetry breaking 
parameters would be in order.
\footnote{The various flavor violating processes have been
examined in the phenomenological approach of the
democratic mass matrices\cite{hamaguchi}.}
It has been known \cite{ross,kobayashi} that the PR fixed point
induces peculiar relations among soft parameters also.
Above all A-parameters are found to be aligned in the present
scenarios with the same dynamics for the Yukawa coupling alignment.
The soft scalar masses for quarks and leptons become flavor
universal at low energy due to large corrections by the democratic 
Yukawa couplings as well as the gauge couplings.
These properties are very desirable for the flavor problem 
in supersymmetric extensions.

However the strong gauge dynamics enhances the soft scalar masses
and also the A-parameters to be comparable with the strongly
coupled gaugino mass.
Therefore the gaugino mass in the strongly coupled sector must
be much smaller than the MSSM gaugino masses.
In this respect, extension to extra dimensions will be
considerable, since FCNC and also CP problems may be solved
or ameliorated without assuming such a special case \cite{extradim}.
Otherwise it would be more natural for our scenarios
that supersymmetry breakings are mediated below the 
GUT scale {\it e.g.} by gauge mediation mechanism.
Further studies of the supersymmetry breaking parameters 
in our models will be discussed elsewhere.

\section*{Acknowledgements}
T.~K. is supported in part by the Grants-in-Aid for Scientific Research
(No. 16028211) and the Grant-in-Aid for the 21st Century COE
``The Center for Diversity and Universality in Physics"
from the Ministry of Education, Science, Sports and Culture, Japan.
H.~T. are supported in part by the Grants-in-Aid for Scientific Research
(No. 13135210) from the Ministry of Education, Science, Sports 
and Culture, Japan.


\begin{thebibliography}{99}
\bibitem{SK} Super-Kamiokande Collaboration,
Y.~Fukuda {\it et al.}, Phys.~Lett.~B {\bf 467}, 185 (1999);
S.~Fukuda {\it et al.}, Phys.~Rev.~Lett. {\bf 85}, 3999 (2000);
Phys.~Rev.~Lett. {\bf 86}, 5651 (2001);
Phys.~Rev.~Lett. {\bf 86}, 5656 (2001).

\bibitem{SNO} SNO Collaboration, Q.~R.~Ahmad {\it et al.},
Phys.~Rev.~Lett. {\bf 87}, 071301 (2001); {\bf 89}, 011301 (2002);
{\bf 89}, 011302 (2002).

\bibitem{KM} KamLAND Collaboration, K.~Eguchi {\it et al.},
Phys.~Rev.~Lett. {\bf 90}, 021802 (2003).

\bibitem{K2K} K2K Collaboration, M.~H.~Ahn {\it et al.},
Phys.~Rev.~Lett. {\bf 90}, 041801 (2003).

\bibitem{models}
For recent reviews, see for example:
H.~Fritzsch and Z.~Z.~Xing, 
Prog.~Part.~Nucl.~Phys. {\bf 45}, 1 (2000);
S.~F.~King,
Rept.~Prog.~Phys. {\bf 67}, 107 (2004);
G.~Altarelli and F.~Feruglio, hep-ph/0405048.

\bibitem{FN}
C.~D.~Froggatt and H.~B.~Nielsen,
Nucl.~Phys.~B {\bf 147}, 277 (1979).


\bibitem{SU3}
See {\it e.g.}, S.~F.~King and G.~G.~Ross,
Phys.~Lett.~B {\bf 520}, 243 (2001); 
Phys.~Lett.~B {\bf 574}, 239 (2003).

\bibitem{quarksector}
H.~Fritzsch, 
Nucl.~Phys. B {\bf 155}, 189 (1979);
Y.~Koide, 
Phys.~Rev.~D {\bf 28}, 252 (1983); 
Phys.~Rev.~D {\bf 39}, 1391 (1989).

\bibitem{leptonsector}
H.~Fritzsch and Z.~Z.~Xing, 
Phys.~Lett.~B {\bf 372}, 265 (1996); 
Phys.~Lett.~B {\bf 440}, 313 (1998); 
Phys.~Rev.~D {\bf 61}, 073016 (2000);
Phys.~Lett.~B {\bf 598}, 237 (2004).

\bibitem{S3}
M.~Tanimoto, Phys.~Lett.~B {\bf 483}, 417 (2000); 
M.~Fukugita, M.~Tanimoto and T.~Yanagida,
Phys.~Rev.~D {\bf 57}, 4429 (1998).

\bibitem{GUT}
M.~Fukugita, M.~Tanimoto and T.~Yanagida,
Phys.~Rev.~D {\bf 57}, 4429 (1998). 

\bibitem{S3improve}    
M.~Tanimoto, T.~Watari, T.~Yanagida, 
Phys.~Lett. {\bf B461} (1999) 345;
E.~Kh.~Akhmedov, G.~C.~Branco, F.~R.~Joaquim and J.~I.~Silva-Marcos,
Phys.~Lett. {\bf B498} (2001) 237;
G.~C.~Branco and J.~I.~Silva-Marcos, 
Phys.~Lett. {\bf B526} (2002) 104;
T.~Watari and T.~Yanagida, Phys.~Lett. {\bf B544} (2002) 167;
Q.~Shafi and Z.~Tavartkiladze, hep-ph/0401235.

\bibitem{extended}
G.~C.~Branco and  J.~I.~Silva-Marcos,
Phys.~Lett.~B {\bf 359}, 166 (1995);
G.~C.~Branco, M.~N.~Rebelo and J.~I.~Silva-Marcos,
Phys.~Lett.~B {\bf 428}, 136 (1998);
Phys.~Rev.~D {\bf 62}, 073004 (2000);
E.~K.~Akhmedov, G.~C.~Branco, F.~R.~Joaquim and 
J.~I.~Silva-Marcos,
Phys.~Lett.~B {\bf 498}, 237 (2001).

\bibitem{NS}
A.~E.~Nelson and M.~J.~Strassler,
JHEP {\bf 0009}, 030 (2000); JHEP {\bf 0207}, 021 (2002);
T.~Kobayashi and H.~Terao, 
Phys.~Rev.~D {\bf 64}, 075003 (2001);
T.~Kobayashi, H.~Nakano and H.~Terao
Phys.~Rev.~D {\bf 65}, 015006 (2002);
T.~Kobayashi, H.~Nakano, T.~Noguchi and H.~Terao
Phys.~Rev.~D {\bf 66}, 095011 (2002).

\bibitem{powerlaw}
M.~Bando, T.~Kobayashi, T.~Noguchi, K.~Yoshioka,
Phys.~Lett.~B {\bf 480}, 187 (2000); 
Phys.~Rev.~D {\bf 63}, 113017 (2001).

\bibitem{PR}
B.~Pendelton and G.~G.~Ross, 
Phys.~Lett.~B {\bf 98} 291 (1981).

\bibitem{Ross}
G.~G.~Ross,
Phys.~Lett.~B {\bf 364}, 216 (1995);
B.~C.~Allanach and S.~F.~King,
Phys.~Lett.~B {\bf 407}, 124 (1997).

\bibitem{AK}
S.~A.~Abel and S.~F.~King,
%``Flavour democracy in strong unification,''
Phys.~Lett.~B {\bf 435}, 73 (1998).

\bibitem{DT}
R.~Barbieri, G.~Dvali and  A.~Strumia,
Phys.~Lett.~B {\bf 333}, 79 (1994);
T.~Yanagida,
Phys.~Lett.~B {\bf 344}, 211 (1995);
T. Hotta, Izawa K.-I., T. Yanagida,
Phys.~Rev.~D {\bf 53},  3913 (1996).

\bibitem{witten}
E.~Witten, hep-ph/0201018;
M.~Dine, Y.~Nir and Y.~Shadmi,
Phys.~Rev.~D {\bf 66} 115001 (2002).

\bibitem{quasifp}
C.~T.~Hill,
Phys.~Rev.~D {\bf 24}, 691 (1981).

\bibitem{nonfree}
M.~Bando, J.~Sato and  K.~Yoshioka, 
Prog.~Theor.~Phys. {\bf 98}, 169 (1997);
Prog.~Theor.~Phys. {\bf 100}, 797 (1998).

\bibitem{deconstruction}
N.~Arkani-Hamed, A.~G.~Cohen and  H.~Georgi,
Phys.~Rev.~Lett. {\bf 86}, 4757 (2001);
Phys.~Lett.~B {\bf 513}, 232 (2001);
C.~T.~Hill, S.~Pokorski and J.~Wang,
%``Gauge invariant effective Lagrangian for Kaluza-Klein modes,''
Phys.~Rev.~D {\bf 64}, 105005 (2001).

\bibitem{hall}
L.~J.~Hall, R.~Rattazzi and U.~Sarid, 
Phys.~Rev.~D {\bf 50}, 7048 (1994);
R.~Hempfling, 
Phys.~Rev.~D {\bf 49}, 6168 (1994).

\bibitem{anarchy}
L.~Hall, H.~Murayama and N.~Weiner,
Phys.~Rev.~Lett. {\bf 84}, 2572 (2000);
N.~Haba and H.~Murayama,
Phys.~Rev.~D {\bf 63}, 053010 (2001);
A.~de Gouvea and  H.~Murayama,
Phys.~Lett.~B {\bf 573}, 94 (2003). 

\bibitem{hamaguchi}
K.~Hamaguchi, M.~Kakizaki and M.~Yamaguchi, 
Phys.~Rev. D {\bf 68} 056007 (2003).

\bibitem{ross}
M.~Lanzagorta and G.~G.~Ross,
Phys.\ Lett.\ B {\bf 364}, 163 (1995);
S.~F.~King and G.~G.~Ross,
Nucl.\ Phys.\ B {\bf 530}, 3 (1998).

\bibitem{kobayashi}
T.~Kobayashi and K.~Yoshioka,
Phys.~Rev. D {\bf 62}, 115003 (2000);
T.~Kobayashi and H.~Terao,
Phys.~Lett. B {\bf 489}, 233 (2000).

\bibitem{extradim}
J.~Kubo and  H.~Terao,
Phys.~Rev. D {\bf 66}, 116003 (2002);
Y.~Kajiyama, J.~Kubo and H.~Terao,
Phys.~Rev. D {\bf 69}, 116006 (2004);
K-Y.~Choi, Y.~Kajiyama, H.~M.~Lee and  J.~Kubo,
Phys.~Rev. D {\bf 70}, 055004 (2004).


\end{thebibliography}
\end{document}